\documentclass[smallextended]{svjour3} 

\usepackage[utf8]{inputenc}
\usepackage{hyperref}
\usepackage{graphicx}
\usepackage{makecell}
\usepackage{pdflscape}
\usepackage{dirtytalk}
\usepackage{csquotes}
\usepackage{framed}
\newcommand{\highlight}[1]{\textbf{Implication: }{\emph{#1}}}
\usepackage{anyfontsize}
\usepackage{adjustbox}
\usepackage{geometry}
\usepackage{pdflscape}
\usepackage{tabulary}
\usepackage{caption}
\usepackage[dvipsnames]{xcolor}
\usepackage{enumitem,amssymb}
\usepackage{float}
\newlist{todolist}{itemize}{2}
\setlist[todolist]{label=$\square$}

\newenvironment{boxed}
    {\begin{center}
    \begin{tabular}{|p{0.8\textwidth}|}
    \hline
    }
    { 
    \\\hline
    \end{tabular} 
    \end{center}
    }

\begin{document}

\title{Factors and actors leading to the adoption of a JavaScript framework}
\titlerunning{Factors and actors leading to the adoption of a JavaScript framework}        

\author{Amantia Pano \and Daniel Graziotin \and Pekka Abrahamsson}


\institute{Amantia Pano \at
              Faculty of Computer Science, Free University of Bozen-Bolzano, Italy \\
              \email{amantia.pano@unibz.it}           
           \and
           Daniel Graziotin \at
              Institute of Software Technology, University of Stuttgart, Germany \\
              \email{daniel.graziotin@informatik.uni-stuttgart.de}           
           \and
           Pekka Abrahamsson \at
              Faculty of Information Technology, University of Jyv{\"a}skyl{\"a}, Jyv{\"a}skyl{\"a}, Finland \\
              \email{pekka.abrahamsson@jyu.fi}           
}

\date{Received: date / Accepted: date}

\maketitle

\begin{abstract}
The increasing popularity of JavaScript has led to a variety of JavaScript frameworks that aim to help developers to address programming tasks. However, the number of JavaScript frameworks has risen rapidly to thousands of versions. It is challenging for practitioners to identify the frameworks that best fit their needs and to develop new ones which fit such needs. Furthermore, there is a lack of knowledge regarding what drives developers towards the choice. This paper explores the factors and actors that lead to the choice of a JavaScript framework. We conducted a qualitative interpretive study of semi-structured interviews. We interviewed 18 decision makers regarding the JavaScript framework selection, up to reaching theoretical saturation. Through coding  the interview responses, we offer a model of desirable JavaScript framework adoption factors. The factors are grouped into categories that are derived via the Unified Theory of Acceptance and Use of Technology. The factors are performance expectancy (performance, size), effort expectancy (automatization, learnability, complexity, understandability), social influence (competitor analysis, collegial advice, community size, community responsiveness), facilitating conditions (suitability, updates, modularity, isolation, extensibility), and price value. A combination of four actors, which are customer, developer, team, and team leader, leads to the choice. Our model contributes to the body of knowledge related to the adoption of technology by software engineers. As a practical implication, our model is useful for decision makers when evaluating JavaScript frameworks, as well as for developers for producing desirable frameworks.
\keywords{
JavaScript \and programming frameworks \and decision making \and technology adoption \and human aspects of software development \and qualitative research \and interpretivism 
}
\end{abstract}

\section{Introduction}

JavaScript has become one of the fundamental programming languages for the Web, and it has been in the top 10 most popular programming languages since the last decade~\cite{Tiobeb}. According to a survey of W3Techs, 88.2\% of one billion websites analyzed rely on this technology~\cite{W3Techsd}. The increasing presence and demand of web users has led to the increasing complexity of web-based software.

Complexity in developing new applications threatens the continuous evolution of programming languages. One of the goals of software engineering is the creation of reusable code~\cite{Fayad1997c}. This goal was initially accomplished by using modules. Yet, modules are limited to resolving small programming tasks. Programmers develop libraries containing pre-written JavaScript code so as to ease and shorten the development of projects. Libraries support the functional and aesthetic aspects of web-centric development. Libraries can be grouped according to their functionalities or they can be part of an application skeleton that is commonly referred to as a framework. JavaScript frameworks comprise a set of utilities, functions, and high-level abstractions, which have been tested in different platforms and browsers. Frameworks serve different purposes, including visual design, charting and dashboards, animation, drag and drop, and event handling.

The rising number of applications relying on JavaScript has resulted in an increased activity of open source communities. Several new JavaScript frameworks have been released and others have been extended with new functionalities~\cite{Gizas2012d}. GitHub~\cite{github} has recently started offering advanced search capabilities on programming languages and various statistics. JavaScript is currently the top programming language with more than 440,000 projects having at least one GitHub star~\cite{githubstarredrepos}. Practitioners have noted this proliferation of frameworks and cope with the difficulties arising from it with running jokes, e.g.,~\cite{Craver2017}.

New but also experienced developers face difficulties in ``making sense out of the jungle of JavaScript frameworks''~\cite{Graziotin2013bd}, especially in defining the most suitable characteristics for a favorable framework to use for their projects~\cite{Ocariza2013b,Graziotin2013bd,Walsh2007c,Reyes2010b,Bennett2007c}. Gizas et al.~\cite{Gizas2012d} have argued that for web developers, it is important to select a JavaScript framework that suits their needs and provides high quality and high performance of the code. They opted for a technical evaluation of JavaScript frameworks using performance benchmarks. While we do not deny the importance of the technical aspects of performance and software quality, there is more to it.  Gizas et al.~\cite{Gizas2012d} have admitted that crucial factors could influence the adoption, such as maintainability and active community support. The previous statement is, however, anecdotal.

Graziotin and Abrahamsson~\cite{Graziotin2013bd} took hold of the previous statement and noted that the limited research on the factors influencing the adoption of a JavaScript framework have insofar focused on performance and quality only. They suggested investigating the developers' perceptions and experience when it comes to adoption. Developers' experience and opinions are shared in web communities or blogs. For example, Walsh~\cite{Walsh2007c} has shared his viewpoint that, in addition to how fast and big the code is, practitioners should also observe the availability of documentation, the community involvement, and the general adoption of the framework. Reyes~\cite{Reyes2010b} has added that practitioners should assess the maturity of the framework, the provided functionality, and the update frequency. Again, these statements come from experience and are valuable; however, they come from single sources and are anecdotal.

We share the previous concern regarding a lack of understanding on the factors that lead to the adoption of a JavaScript framework. An understanding of the adoption factors of a JavaScript framework will lead to the creation of new frameworks that answer the needs of developers, as well as the possibility of the meaningful categorization of the existing frameworks for supporting a choice.

Our paper explores the factors and actors leading to the adoption of a JavaScript framework as requested by previous research, e.g.,~\cite{Cunha:2016kw,Graziotin2013bd}. The research question that we aim to answer is \textit{Which factors and actors lead to the adoption of a JavaScript framework?}. We report a qualitative interpretive study of semi-structured interviews during which we interviewed 18 participants. We recruited the participants using various channels, and we refined our participation criteria while we gathered data. Participants were either expert developers, decision makers in their companies, or entrepreneurs, or were at least able to motivate the adoption of a JavaScript framework.

The analysis of the qualitative data revealed:

\begin{enumerate} 
\item A model of features that are desirable in a JavaScript framework---which are grouped into the areas of the Unified Theory of Acceptance and Use of Technology (UTAUT)~\cite{venkatesh2003}, namely performance expectancy, effort expectancy, social influence, facilitating conditions, and price value~\cite{venkatesh2012}.
\item A representation of the decision makers involved in the framework selection---namely customer, developer, team, and team leader---which is based on the model. 
\end{enumerate}

The remainder of this article is divided into five sections. Section~\ref{sec:related_work} reports our literature review of the research in software and web engineering and practitioners' voices. Section~\ref{sec:methodology} reports our methodology, in terms of the epistemological stance, participants' recruitment, interview design, and data analysis. Section~\ref{sec:results} provides the resulting model. Each item is demonstrated by one or more interview snippets that brought us to elicit the particular category or factor. In section~\ref{sec:discussion}, we compare our model with the existing literature, we suggest several implications, and we provide the limitations of this study. Section~\ref{sec:conclusions} concludes the paper.

\section{Related work}
\label{sec:related_work}

In this section, we present the works that we identified as relevant to our study. First, we summarize the related work from the scientific literature. Then, we provide a selection of relevant practitioners' viewpoints. The reason for including the grayer literature by developers and leaders is that they produced more related material---in technical blogs, reports, and dedicated websites---than what has been published in the scholarly literature.

\subsection{Scientific literature}

The software engineering body of knowledge has shown a limited understanding of the question we address in this paper. 

Gizas et al.~\cite{Gizas2012d} stated that we must consider three aspects when comparing JavaScript frameworks: quality, performance and validation tests on the frameworks. The authors compared the frameworks ExtJS~\cite{extjs}, Dojo~\cite{dojo}, jQuery~\cite{jquery}, MooTools~\cite{mootools}, Prototype~\cite{prototypejs}, and YUI~\cite{yui}. During the research they identified four different run-time versions for each language: (1) Base Version (uncompressed) which includes the core functions only; (2) Compact Version, i.e., the base version with no comments and blank lines; (3) Full Compact Versions, with all the available functions in a compact format; (4) Development Kits, with all the development tools to satisfy more advanced developer needs. Gizas et al.~\cite{Gizas2012d} defined common functions to use for the evaluation: Document Object Model (DOM)\footnote{DOM is an application programming interface (API) set and a model that defines the logical structure of web documents and the way a document is accessed and manipulated.~\cite{W3DOM}} manipulation, selectors, Asynchronous JavaScript and XML (AJAX)\footnote{AJAX is a set of web development practices and techniques for enabling asynchronous web applications on the client-side~\cite{garrett2005ajax}} functionalities, basic elements of forms, functions for base event handling and functions for compatibility support and loading utilities. Gizas et al.~\cite{Gizas2012d} measured quality of the classic software metrics of measuring size (e.g., lines of code), complexity (e.g., McCabe's Cyclomatic Complexity~\cite{mccabe1976complexity}), and maintainability (e.g., Maintainability Index). The tools used for the quality tests were: JSmeter, CLOC, and Understand. The validation tests were conducted using the Yasca software utility, in combination with JavaScript Lint. The authors employed different browsers and operating systems for the performance tests, which they analyzed with the SlickSpeed Selectors test framework~\cite{Gizas2012d}.

Misra and Cafer~\cite{Misra2012e} introduced a complexity metric for JavaScript, called JavaScript Cognitive Complexity Measure (JCCM). The metric is intended to access the design quality of scripts. The JCCM metric formula is calculated using (1) the size in terms of lines of code containing variables and operators, (2) the number of arbitrarily named distinct variables (ANDV), (3) the number of meaningfully named variables (MNV), (4) the cognitive weights of basic control structures (BCS) such as loops, recursions, etc., and (5) the number of operators. The authors analyzed thirty JavaScript files from the Web. The authors showed that the metric converges to the metrics of Cyclomatic Complexity~\cite{mccabe1976complexity}, Logical Lines of Code, and Halstead metrics~\cite{Halstead1977c}. The paper does not report any validation from the practitioners' point of view.

Graziotin and Abrahamsson~\cite{Graziotin2013bd} conducted a pilot qualitative study with experienced web developers for an exploratory model for the comparison of JavaScript frameworks. The model suggests that there are three criteria that drive practitioners towards the choice of a JavaScript framework, namely documentation, community, and the pragmatics of a JavaScript framework. Additionally, the findings suggest that it is important for practitioners to have software metrics applied to a single software project implemented using different JavaScript frameworks, in order to have meaningful comparisons. The pilot study was conducted on a very limited sample of practitioners (four), but it lays down the basic motivations and foundations of the present work.

In addition to the previous papers dealing with the adoption of JavaScript frameworks, we found two related works on the factors leading to the adoption of programming language add-ons among developers. The first tested a theory on the factors leading to the adoption of software components. The second tested a theory on the factors leading to the adoption of a programming framework.

Stefi~\cite{stefi2015} investigated the developers' intention to adopt software components. Stefi collected 142 answers from a questionnaire distributed to software developers. The construction of the questionnaire was based on the Unified Theory of Acceptance and Use of Technology (UTAUT)~\cite{venkatesh2003}, which we also adopted for the present study and will explain in Section~\ref{sec:data_analysis}, with the addition of items representing the not-invented-here bias and mindfulness to account for individual differences. The participants rated items that were related to a post-adoption rationale. The results show that performance expectancy, social influence, and the not-invented-here bias play the biggest role in the intention of adopting software components.

Polan\v{c}i\v{c} et al.~\cite{polancic2010} conducted an online survey of 389 software developers for examining the major drivers towards the adoption of a framework and its success. The authors framed the creation of the survey items with the Technology Acceptance Model (TAM)~\cite{r02166}, a predecessor of UTAUT that explains how users arrive at accepting any technology through the perceived usefulness and ease-of-use. The survey asked the respondents to identify a framework that they were either developing or using. Then, participants rated Likert and semantic differential scale items also related to TAM. That is, participants quantitatively explained a post-adoption decision. The authors found that a successful adoption of a framework is mostly driven by a continuous intention to use the framework and its perceived usefulness. Additionally, the authors empirically demonstrated that TAM can be applied to the post-adoption research of frameworks, and that the post-adoption TAM constructs anticipate the success of a framework. The investigated programming languages were mostly Java, and the .NET framework, and JavaScript is not mentioned.

\subsection{Practitioner viewpoints}

To further investigate the overall knowledge in the field, we selected some insightful pieces in the gray literature of developers' personal experience and the advice they shared with others, and the most prominent technical websites specialized in our topic of interest.

According to Bennet~\cite{Bennett2007c}, there are four parameters that a JavaScript framework must ensure in order to be considered as a meaningful choice. The parameters are (1) a normalized event model, including attaching and removing listeners, automatic scope correction and access to the event itself, (2) a normalized wrapper around XMLHttpRequest\footnote{XMLHttpRequest is a specification for an API that provides client-side functionality for transferring data between a client and a server.~\cite{W3XMLHttpRequest}}, including the ability to specify callbacks to fire when the request finishes, (3) a set of normalized utilities for working with the DOM, especially for getting or setting styles and managing class names, and (4) facilities for creating visual animations, whether predefined or custom, which handle cross-browser timing issues. Bennet also suggested  that a JavaScript framework should present the scripting language as it is and not make it similar to other languages, and that the DOM methods should be used only if wrapping is necessary for cross-browser compatibility. Also, the name of objects inside a JavaScript library or framework should be meaningful and possibly be inherent to the functionalities that the library or framework provides, and that the documentation is very important for easily understanding a framework and extending its functionalities. 

Reyes~\cite{Reyes2010b} oriented towards an analysis of the maturity of a framework by retrieving information on how long it has been developed and how often new updates are released. It is important to Reyes that the community behind a benchmark consists of experienced developers. He took into consideration the requirements of the framework with respect to the requirements of the web page. Programmers should focus on the usability of the framework. Reyes suggested checking if the framework needs continuous updates and if the programmers are required to know the libraries and frameworks very well. The analysis also focused on the documentation, which is crucial for satisfying information needs.

Walsh~\cite{Walsh2007c} advised analyzing the speed of a JavaScript framework by using the MooTools Slick Speed test~\cite{mootoolsspeed}. The tool evaluates the speed and validity of selectors of Cascading Style Sheets  (CSS) version 3\footnote{CSS is a set of style rules that apply visual properties to elements of a web document~\cite{W3CSS}} for MooTools 1.2~\cite{mootools}, MooTools 1.3.1 ~\cite{mootools},  jQuery 1.5.1~\cite{jquery}, Prototype 1.7~\cite{prototypejs}, YUI 2.8.2 Selector~\cite{yui}, and Dojo 1.5~\cite{dojo}. A JavaScript framework should not be \textit{heavy} in terms of storage as it could grow while features are added or extended. The author considered modularity as a crucial aspect to check for. The community and its activity are important, as Walsh considered those factors as reflective of quality and interest. As with any other programming language, it is helpful to identify if the needs of the web application will be fulfilled and if implementation is effectively time consuming or not. 

While reviewing the practitioners area, we found three interesting projects aiming to make sense of JavaScript frameworks. 
Two of them terminated their service during the peer review of this paper.

Jster.net~\cite{jster} offers a categorization of 1780 frameworks and libraries\footnote{They were 1573 when we first submitted the present paper.}. The catalog is divided into the following categories: Essential frameworks; User Interface; Multimedia; Graphics; Data; Development; Utilities; and Applications. The entries are ordered within each category in terms of their popularity on GitHub as aggregated by the number of project favorites, forks, and a community-based rating. There appears to be no support for the longevity of the libraries, which might become obsolete over time.

The website designzum.com~\cite{designzum} offered a ranking of JavaScript frameworks based on parameters such as weight (in terms of code), ease of learning and implementing, modularity, and simplicity of wrapping up with new features. The list focused on two keywords: \textit{lightweight} and \textit{simple to learn}. The website jsdb.io~\cite{jsdb} ranked the JavaScript frameworks by exploiting four parameters coming from the GitHub~\cite{github} repositories of the frameworks. The metrics used were composed of (1) stars, or the number of GitHub~\cite{github} users observing the project, (2) the average length of time between code commits, (3) the number of contributors in the last 100 commits, (4) the number of forks made for a given project.

The review of the literature and of the practitioner's sites has suggested that scientific research has developed metrics that can be applied for measurements, and it also introduces a model which is based on subjective criteria arising from qualitative research on the practitioners' feedback. Practitioners, however, appear to value more practical features of a JavaScript framework or library that are important for developers. Niche technical websites have adopted miscellaneous, non-validated variables to rank or list JavaScript frameworks. 

This abundance of non-scientifically validated parameters has led us to research a model which gathers the empirical and theoretical knowledge for identifying the relevant variables that guide practitioners towards the choice of a JavaScript framework.

\section{Methodology}
\label{sec:methodology}

In this section, we describe our chosen methodology. First, we characterize the study in terms of our worldview. Secondly, we clarify variance-based and process-based research, so that we can characterize the output of our study. Thirdly, we provide a description of the participants' recruitment process and criteria. Finally, we describe how we analyzed the data obtained from the interviews.

\subsection{Interpretive research}

Our aim is to deliver new knowledge concerning the features that decision makers weight as important when selecting a JavaScript framework. We research human behavior based on the perceptions that the individuals offer us when asked on their adoption of a framework. We recognize that our analysis of the participants' perceptions will be further interpreted by us. This is why we conduct interpretive research, which is frequently adopted when producing theories and models for explaining phenomena~\cite{klein1999}. Interpretivism is established in information systems research~\cite{Walsham2006}, but it is still emerging in software engineering research. For this reason, we describe it here briefly.

Interpretive research derives from interpretivism, often interchanged with social constructivism~\cite{easterbrook2008,klein1999}, and it is defined by Geertz~\cite{geertz1973} as ``really our own constructions of other people’s constructions of what they and their compatriots are up to'' (p. 9). The assumption of this worldview is that individuals (participants and researchers) seek an understanding of the world they live in, and develop subjective meanings of their experience toward objects, things, and phenomena~\cite{Creswell2009}.

Given the novelty of the research question and the lack of knowledge towards its answer, we opted to design a qualitative analysis using semi-structured interviews with open-ended questions to be executed and analyzed with the inspiration of grounded theory~\cite{Langley1999}.

\subsection{Variance based and process based research} We would like to clarify that the present work, while exploring the adoption process, is not process-based research but variance-based.

Variance-based models and theories provide explanations for phenomena in terms of relationships among dependent and independent variables~\cite{Langley1999,Mohr1982}. In variance theory, the precursor is both a necessary and sufficient condition to explain an outcome,  and the time ordering among the independent variables is immaterial~\cite{Pfeffer1983,Mohr1982}. Based on our experience with the literature, the majority of software engineering research is variance-based, yet process-based studies are rising.

Process-based research development has its roots in studies attempting to explain how and why organizations change~\cite{VanDeVen1995}. Nowadays, process-based research activities attempt to understand how \emph{things} evolve over time and why they evolve in the way we observe~\cite{Langley1999}. Process data consist mainly of \emph{stories}---which are implemented using several different strategies---about what happened during the observation of events, activities, choices, and the people performing them over time~\cite{Langley1999}. 

The output of the present work, while investigating the adoption process, is of variance type. That is, we present a model of factors that influence the choice of a JavaScript framework but we do not offer a representation of the adoption process itself. One of the implications of this choice is that we do not offer any representation of the decision-making process per se, but the static representation of the factors that lead to the adoption of a framework.

\subsection{Participants recruitment}

We planned the selection and recruitment of the participants so as to identify individuals who could enrich the research outcome through sharing their experiences and thoughts. We followed a judgmental sample strategy, meaning that we followed a non-probability sampling technique and we purposely looked for experts in order to analyze their experience. Criteria for participation was to be either expert developers, decision makers in their companies, or entrepreneurs, or were at least able to motivate their JavaScript frameworks' adoption. As we analyzed the data, we performed small refinements to the criteria (e.g., looking for participants with in-house developed frameworks instead of adopting an open-source one) until reaching theoretical saturation as suggested by Strauss and Corbin~\cite{Strauss2008b}. We also aimed to obtain a sample size in the range of the suggested number of participants in qualitative studies~\cite{Guest2006c}.

The recruitment process started by searching for developers who were working actively in the web development field and who were competent, proficient and experts in the domain.

The time it takes to be proficient in a programming language strongly depends on the programmer's acquired knowledge~\cite{Robins2003c}. The model proposed by Dreyfus and Dreyfus~\cite{Dreyfus1987b}, which describes skills acquisition, proposes these stages: novice, advanced beginner, competent, proficient, and expert. This model states that novice and advanced beginners have no recognition of the relevance of the task they are assigned to, while those who are competent, proficient and expert do. The first two levels of skills apply to those individuals who are not autonomous and are not able to deal with the resolution of complex situations. We expected the participants to belong to the competent, proficient, and expert level of skills in the field. We expected the developers to be decision makers and to have a broader view on the technologies used for web development and particularly JavaScript. Furthermore, we expected them to have a historical knowledge of the frameworks and their evolution over time. 

Developers behind JavaScript frameworks tend to be part of open source communities. They share their knowledge on forums, blogs, web pages or online magazines, so the recruitment phase started on the Web. 

The recruitment process employed social networks as a primary source for finding candidates. We identified the \textit{hashtag} (\textit{\#}) symbol and tool as a key to filtering the information spread over the Web. The hashtag is a form of meta data tag used in social networks such as Twitter~\cite{twitter}, Facebook~\cite{facebook}, and Google+~\cite{googleplus}. Hashtags allow social network users to post information on a specific topic by using the same tag. This allows people who search for a specific keyword to have all related posts grouped in the view. 

We structured the research by keywords into three approaches, all of which employed different social networks. The first approach focused on the widely adopted social networks Twitter~\cite{twitter}, Facebook~\cite{facebook}, and Google+~\cite{googleplus}. We inspected these three social networks using the following hashtags: \textit{\#javascript}, \textit{\#javascript \#framework}, and \textit{\#web \#development}. 

The second approach was to research more specific figures based on the participants' skills and knowledge in a more formal platform such as LinkedIn~\cite{linkedin}. 

The third approach consisted of researching within the communities related to JavaScript within the social coding platform GitHub~\cite{github}.

We reviewed the initial query results every 4 hours for a week. The research results returned posts by developers, help requests from novices in the field, talks from journals and posts for advertising services. We focused on the posts of developers. We inspected their profiles to see the description of their working area and skills. In those cases where a website was included in the profile description, we inspected the website for any additional information on the candidate that could be helpful to identify his current work and past experience. In some occurrences the reference websites were blogs maintained by the candidates. We prepared an invitation text for the recruitment, which was adapted depending on whether more detailed information on the candidate could be retrieved.

E-mail recruitment offered a brief explanation of where the contact was retrieved from and the purpose of the research. We also specified the time commitment expected of the interviewee. We obtained either a verbal or written consent prior to recording the interview. The e-mail specified that participation was on a voluntary basis and that to drop out at any point was possible without consequences. Participants were informed that we would not collect data for third parties, that the data would remain with us and the transcripts would not be published but only the research findings supported by anonymous quotes. We used participant codes to identify candidates so that the responses could not be traced back to the participant's confidential data. It was mentioned that no detailed information concerning the company they were working for would be asked for or revealed in the final report. We ended with our contact details, which could be used in case of interest.

The second approach focused on a professional social network, which is LinkedIn~\cite{linkedin}. Here, professionals complete their profile descriptions by adding skills and certifications. The recruitment in this approach was based upon the skills. Candidates were recruited by sending them direct messages. 

The third approach was to get into the heart of JavaScript framework design, by analyzing GitHub~\cite{github} repositories. Some JavaScript framework communities were chosen based on popularity such as AngularJS~\cite{angularjs}, jQuery~\cite{jquery}, NodeJS~\cite{nodejs}, etc. GitHub~\cite{github} allows the research of the community for the most trending repositories and developers of the month for any of the languages in their repositories. We selected the candidates appearing in this view for the month of July 2014. We contacted them if they provided any e-mail or company contact information.

\subsection{Interview design}

The qualitative research interview, especially when undertaken through an interpretive worldview, seeks to describe the world and the meaning of specific phenomena based on the perception of the participants~\cite{Creswell2009}. The information retrieved during the interview stems from the experiences of the participant. The interviewer can deepen the understanding of the responses by asking additional questions as the process is interactive~\cite{McNamara1999b}.

For the research design, we chose to employ a semi-structured interview approach, and we formulated open-ended questions. This choice allowed us to briefly enter the subject of the research and to discover new information if the persons were less eloquent than expected. The starting questions, which are the results of a refinement process as the interviews were proceeding, are available in the appendix (Section~\ref{appendix}).
Over time, fifteen practitioners agreed to participate in the interview and three of them participated via e-mail. The interviews were conducted on the phone, personally, via Skype and Google Hangouts, and via e-mail.
Interviewees were asked to respond with their availability and a time schedule appropriate for them for the interview. 

We expected the duration of the interviews to be within five to ten minutes. We found it to be an acceptable length as the candidates were participating during their free time. We informed them in advance that the time commitment would not impact significantly on their time schedule.

\subsection{Data analysis}
\label{sec:data_analysis}

We transcribed the interviews immediately after they were concluded. 
Then, we alternated phases of data analysis with refinements of the interview skeleton.
For analyzing the data, we developed codes, which are words or phrases that summarize the content of a data segment, be it a word, sentence, or paragraph~\cite{Strauss2008b}.
Extracts of the responses can also be a code and we refer to those as in vivo codes~\cite{Strauss2008b}.

We analyzed the data with QSR International's NVivo 10 software tool~\cite{nvivo}.
Participants' recruitment and data analysis stopped when we agreed to reach theoretical saturation. That is, interviews and analyses were performed iteratively with few participants at a time until the last round of new interviews and analysis did not lead to changes to the resulting model.

We performed three iterative phases of data coding and analysis, namely:

\begin{description}
\item[Phase 1] Characterization and summarization of the interview extracts.
\item[Phase 2] Identification of relationship between codes.
\item[Phase 3] Identification of grouping categories and central categories for the taxonomy. 
\end{description}

We describe these phases with the aid of a coding example, also known as chain of evidence, in Table~\ref{fig:chain-evidence}.

\begin{landscape}
\begin{center}
  \begin{tabulary}{\textwidth}{LLL}
    Interview Extract / Phase 1 & Phase 2  & Phase 3 \\
    \hline
    Interviewer: ``Do you think that the framework you use is mature enough, and why?'' & {\color{OliveGreen}Maturity depends on size of JSF and size of community} & {\color{blue}Factors}-{\color{magenta}Social\textunderscore influence}-{\color{gray}Community\textunderscore size} \\
    
    P01: “Angular is huge {\color{gray}[size of JSF]} and well developed and the community behind it too {\color{gray}[size of community]}”. & {\color{OliveGreen}Community size and time of activity imply maturity} & {\color{blue}Factors}-{\color{magenta}Social\textunderscore influence}-{\color{gray}Community\textunderscore size} \newline {\color{blue}Factors}-{\color{magenta}Social\textunderscore influence}-{\color{gray}Community \textunderscore responsiveness} \\
    
    P08: ``every framework with time {\color{gray}[activity time]} gets mature when many people around the world try their hands in that {\color{gray}[size of community]}''. & & \\

    Interviewer: ``How close are you with the community?'' & & \\
    P01: ``We rely a lot on the community {\color{gray}[frequent contact with community]} and on Internet sites such as Reddit {\color{gray}[feedback from other programmers]}.'' & {\color{OliveGreen}The community responds promptly as bigger the framework is} & {\color{blue}Factors}-{\color{magenta}Social\textunderscore influence}-{\color{gray}Community \textunderscore responsiveness} \\

    P01 [continued]: ``Communities are quite important, we rely on them every time [frequent contact with community] and as much big the community {\color{gray}[size of community]} as faster you receive responses from them {\color{gray}[fast responsiveness]}'' & {\color{OliveGreen}Questions are replied quickly in a big community} & {\color{blue}Factors}-{\color{magenta}Social\textunderscore influence}-{\color{gray}Community\textunderscore size} \newline {\color{blue}Factors}-{\color{magenta}Social\textunderscore influence}-{\color{gray}Community \textunderscore responsiveness}\\
  \end{tabulary}  
  \captionof{table}{
      Working example of the coding process. \\
      Color coding:
      {\color{gray}[codes for characterization]};
      {\color{OliveGreen} codes for relationships}; 
      {\color{blue} codes for overarching categories};
      {\color{magenta} codes for UTAUT categories};
  }
\label{fig:chain-evidence}
\end{center}
\end{landscape}

Table~\ref{fig:chain-evidence} provides three columns. 
In the first column (Phase 1), we report an interview extract and the corresponding code in square brackets that characterize the various parts of the interview extract. 
Participant P01 used the adjective \say{huge} referring to both the community size and JavaScript framework size, while P08 referred to community size in terms of \say{many people around the world try their hand in that}. These were the characterization and summarization codes we identified in Phase 1. The phase produced 76 codes including labels and in vivo codes.

In the second column (Phase 2), we provide commonalities and relationships between the participants and the codes of Phase 1. There, we connected the codes related to community size, JavaScript framework size, and time of activity with a relationship implying framework maturity. We also connected codes about community feedback with the size of a community. This phase aided us in developing the implications that we summarize in Section~\ref{sec:implications}.

In the third column (Phase 3), we show the final grouping categories of the codes. \textit{Community size} was grouped into the Unified Theory of Acceptance and Use of Technology (UTAUT) category of \textit{social influence} (explained next), which in turn is part of the \textit{factors} category. 
Codes which refer to a similar concept are grouped into a category. 
We identified five categories and twenty subcategories of factors as shown in Figure~\ref{fig:model-factors} and Figure~\ref{fig:model-actors}. 
We also produced two central overarching categories, which are able to cover the entire spectrum of available categories. These central categories are sometimes called themes in the literature and are strictly related to the research question.
The central categories we identified were \textit{Factors} and \textit{Actors} that lead to the adoption of a JavaScript framework.

The coding process that produced the five main categories of Figure~\ref{fig:model-factors} was influenced by the Unified Theory of Acceptance and Use of Technology (UTAUT)~\cite{venkatesh2003}\footnote{We thank an anonymous reviewer for suggesting to look into UTAUT,}. UTAUT was proposed by Venkatesh et al.~\cite{venkatesh2003} to explain the acceptance of technology by users. It is considered as an evolution of the older Technology Acceptance Model (TAM)~\cite{r02166}, as it attempts to reduce its redundancy and isolate the core determinants of user acceptance of technology adoption.''~\cite{vishwanath2016a}. Four main constructs are direct determinants of behavioral intention to adopt a technology, namely performance expectancy, effort expectancy, social influence, facilitating conditions, plus a fifth one, price value (added in a revision of UTAUT~\cite{venkatesh2012})\footnote{Mediating factors to technology adoption are gender, age, experience, and voluntariness of use~\cite{venkatesh2003}. We do not consider these mediating factors as we do not test the theory in this study.}. We explain the constructs when we introduce them in Section~\ref{sec:results}. 

UTAUT is considered to be the leading theory for explaining the adoption of technology~\cite{williams2015}. A recent literature review by Williams et al.~\cite{williams2015} found that UTAUT was employed to explain the adoption of various technologies in several fields (e.g., communication systems, office systems, e-learning technologies). In software engineering and related research fields, UTAUT was employed as intended, by treating developers as technology users, for explaining the adoption tools among developers, such as agile process supporting tools~\cite{hong2011a} and the use of recommended systems in integrated development environments~\cite{gasparic2017}. However, UTAUT (and its predecessor TAM) have also been used for investigating the adoption of programming paradigms such as object-oriented programming~\cite{hardgrave2003a}, learning and acceptance of algorithms~\cite{doukakis2013} and programming languages~\cite{idemudia2016a} and, finally, the developers' intention to adopt existing software components~\cite{stefi2015} and programming language frameworks~\cite{polancic2010}, as we do in our study. All previous studies adopted UTAUT or TAM as a framing component for designing quantitative questionnaires and to study developers' post-adoption intentions. While we also study developers' post-adoption intentions, we adopted UTAUT to help us build the higher-level items of our classification, post data collection.

We inspected each second-level categories of Figure~\ref{fig:model-factors} and discussed if and how they could be part of a UTAUT category. We would create a separate category otherwise. 
We will describe in Section~\ref{sec:results:factors} that we found our categories to be a good fit for all UTAUT original categories with the exception of one (Cost), which is however part of a revision of UTAUT~\cite{venkatesh2012}.

As a final note, the reader might notice a resemblance of our coding phases to those proposed by the grounded theory methodology by Strauss and Corbin in~\cite{strauss1998}, which are called open, axial, and selective coding. We are aware that many different coding strategy have been proposed by several authors \footnote{Stol et al.~\cite{stol2016} have conveniently summarized the major proposals for grounded theory, including coding techniques.}. However, we note that these strategies have evolved over time. As argued by Heath and Cowley~\cite{heath2004}, the history of grounded theory presents several developments, interpretations, and conflicts. Qualitative coding is specifically a part the clash, but a difference between the various proposed coding techniques is debatable and almost negligible, if not uninteresting. Glaser stated that fighting over a true grounded theory methodology is a ``rhetorical wrestle''~\cite{glaser2014}. Researchers, continue Heath and Cowley~\cite{heath2004}, should stop debating about grounded theory, select or develop the method that best suits their cognitive style and research environment, and start doing research. We are of the same stance and define our study as a qualitative interpretive study of semi-structured interviews, which uses elements of grounded theory as proposed by Strauss and Corbin in~\cite{strauss1998}. Hence, our neutral naming of the coding phases. Even if we do not take a stance in the previous matter, we describe our methodology meticulously to allow the evaluation of our study and to ensure replications and extensions.

\section{Results}
\label{sec:results}

This section presents the results of the research. Initially, we provide a classification of the participants based on the judgmental sampling of the recruitment phase. Then, we introduce our classification of factors and actors leading to the adoption of a JavaScript framework. Each category, sub-category, and code is demonstrated by one or more interview quotes.

\subsection{Participants' classification}

We recruited eighteen participants for this study. The participants were located for working purposes in the following countries: Albania, Germany, India, Italy, Netherlands, and the USA. The data was collected via a phone interview for two participants, chat via Google Hangouts for two participants, call via Skype for eleven participants and three participants requested the questions to be sent via e-mail.

The interviewees were classified according to five attributes: (1) years of experience, (2) frameworks used, (3) development focus, (4) company size, and (5) company type. The results are displayed in Table \ref{tbl:sample}. What follows is a description of each attribute used in the classification. 

\begin{landscape}
\begin{table*}
\small
\centering
\begin{tabular}{| r | r | l | l | l |} 
\hline
\thead[r]{ID}  & \thead[r]{Experience\\{years}}   & \thead[l]{frameworks\\used}    & \thead[l]{Development\\focus} & \thead[l]{Company\\Size and Type} \\ \hline
P01 & 4 & \makecell[l]{Angular JS, Gulp JS, Node JS, pure JS} & FB & Small Research \\ \hline
P02 & 4 & \makecell[l]{BackboneJS, ExpressJS, jQuery, NodeJS, pure JS} & F & Independent Worker \\ \hline
P03 & 4 & \makecell[l]{Angular JS, jQuery, RequireJS} & F & Medium Industry \\ \hline
P04 & 5 & \makecell[l]{Own frameworks} & FB & Medium Industry \\ \hline
P05 & 5 & \makecell[l]{Own frameworks} & FB & Small Industry \\ \hline
P06 & 6 & \makecell[l]{Falcon.js, jQuery,  LessJS, MooTools} & F & Large Industry \\ \hline
P07 & 6 & \makecell[l]{jQuery Mobile} & F & Small Industry \\ \hline
P08 & 6 & \makecell[l]{AngularJS, jQuery, UnderscoreJS, VanillaJS} & B & Large Industry \\ \hline
P09 & 6 & \makecell[l]{No frameworks} & F & Independent Worker \\ \hline
P10 & 8 & \makecell[l]{AngularJS, jQuery, ReactJS} & F & Large Industry \\ \hline
P11 & 8 & \makecell[l]{DojoToolkit} & B & Large Industry \\ \hline
P12 & 8 & \makecell[l]{EmberJS, AngularJS, BackboneJS} & FB & Medium Industry \\ \hline
P13 & 9 & \makecell[l]{AngularJS, jQuery, KnockoutJS} & F & Medium Industry \\ \hline
P14 & 10 & \makecell[l]{SpringJS} & B & Medium Research \\ \hline
P15 & 10 & \makecell[l]{jQuery, pure JS} & FB & Small Industry \\ \hline
P16 & 11 & \makecell[l]{Express.js, Node.JS, pure JS} & F & Medium Industry \\ \hline
P17 & 12 & \makecell[l]{jQuery, pure JS} & FB & Medium Industry \\ \hline
P18 & 15 & \makecell[l]{AngularJS, BackboneJS, KnockoutJS, NodeJS, UnderscoreJS} & FB & Large Industry \\ \hline
\end{tabular}
\caption{Classification of participants. F: front-end; B: back-end; FB: front- and back-end}
\label{tbl:sample}
\end{table*}
\end{landscape}

\paragraph{Years of experience} The participants answered the question \say{How long have you been developing for the Web?} The minimum experience was 4 years and the maximum experience in the field was 15 years. 

\paragraph{Framework used} The participants answered the question \say{Do you use any frameworks for helping development?} or \say{Which framework do you use for developing web applications?}. This classification includes three possible assignment strategies: (1) using no frameworks; (2) using own frameworks; or (3) using one or more frameworks, which we listed. The need for this classification arose from the fact that one of the participants stated that no framework was used for developing:

\begin{displayquote}
\say{I don't rely or use any framework, except cases when [the customer] requires it [\ldots] I prefer [pure] JavaScript. It seems more natural to work with the language than library [meant as framework].}---P09\footnote{Our quotations are verbatim excerpts, except for spelling and punctuation corrections. If we applied changes to wording (for comprehensibility), these are indicated by square brackets.}.
\end{displayquote}

Two participants stated as follows: 

\begin{displayquote}
\say{I decided to create it on my own, I created some classes which interact with the underlying system. The framework will be soon available for download. I will distribute it under GPL license, for non commercial and for commercial use.}---P05
\end{displayquote}

\begin{displayquote}
\say{[referring to a former statement of the interview that jQuery according to people is fantastic]  not now, although it helps me greatly expand the idea of what people need [to] make [their] job easier, because i did my own framework}---P04
\end{displayquote}

\paragraph{Development focus} The participants stated whether they were working for the front-end or for the back-end of systems, or for both. Three of the participants were working exclusively for the back-end, six of them stated they were working for both the front and the back-end, while the rest of the participants stated they were working mostly or exclusively for the front-end.

\paragraph{Company size and type}  Two participants stated they were working as freelancers, five were working in large business companies, seven were working for medium-sized companies, and four were working for small companies with a maximum of 5 collaborators. The majority of the developers worked for companies which have IT as their core business while the others work in different areas such as industry (sales, outsourcing, finance) and research. We collected this information in the form of memos during the participant recruitment phase.

\subsection{Factors leading to the adoption of a JavaScript Framework}
\label{sec:results:factors}

Here, we describe our classification of factors leading to the adoption of a JavaScript framework. First, we present the overall classification scheme of the \textit{Factors} central category; then, we describe each discovered category and code which are accompanied by citations of interview extracts. The interview extracts are reported verbatim with respect to what the participants declared. Between some of our codes, we also offer implications that we derived from observations of the relevant data, should we generalize our results. These implications are then listed altogether in Section~\ref{sec:implications}.

\subsection{Factors}

\begin{figure*}
\centering
\includegraphics[width=0.60\textwidth,keepaspectratio]{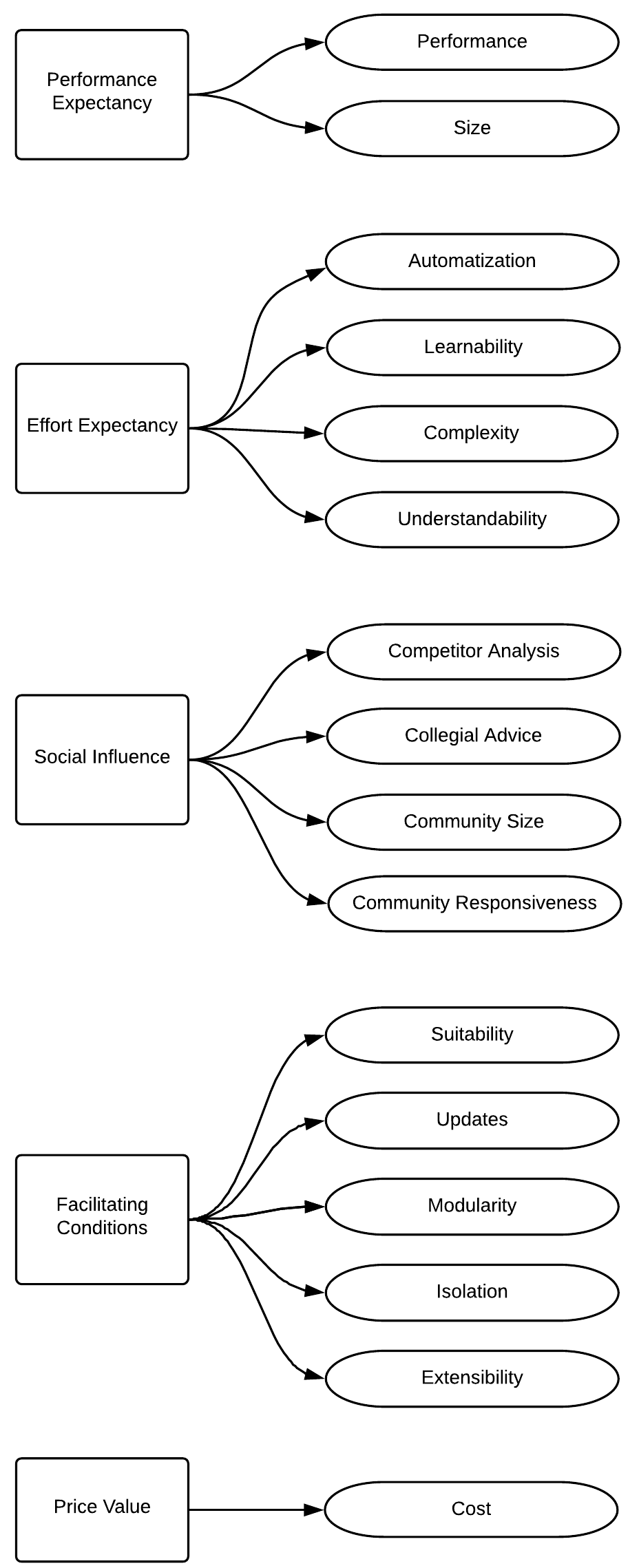}
\caption{Factors leading to the adoption of a JavaScript Framework}\label{fig:model-factors}
\end{figure*}

The \textit{Factors} category groups the attributes of a JavaScript framework according to a set of categories of features. The categories and subcategories that we developed during the coding phase can be observed in Figure~\ref{fig:model-factors}.

\subsubsection{Performance expectancy}

Performance expectancy is defined in UTAUT as the degree to which an individual believes that using the potentially adoptable system will help to attain gains in work or outcome performance~\cite{venkatesh2003}. We rework the original definition to observe the performance gains of the built application instead.\footnote{\label{explanation-performance} One reason for our reworked definition is that, in our interpretation of the data, our participants saw their job-related productivity proxied both by application performance and ease of their daily operations, that is effort expectancy as we describe in the next sub-section. This is in line with the general open debate regarding what developers' productivity actually is (see for example,~\cite{meyer2014,fagerholm2015,kettunen2013}), as well as when applying UTAUT in the context of software engineering, where performance expectancy can be interchanged with effort expectancy~\cite{hong2011a,stefi2015}. We opted to support our participants both ways, and concentrate gains in the product's performance as performance expectancy, while ease of use and programming effort belong to effort expectancy in the next sub-section.}.

\subsubsection*{Performance} The participants explained their concern on the overall performance of the final application and on its size. A drawback was described as when the number of lines of code to build basic functionality is too high or when the payload for including the required functionality is too high in terms of memory and bandwidth. A participant stated the following in this regard:

\begin{displayquote}
\say{[making a decision] may also be related to the requirement that you actually have. Like using React right now seems to promise a lot of performance  benefits [\ldots] to the hardware side}---P10
\end{displayquote}

\begin{displayquote}
\say{Google [PageSpeed Insights] measures all the cache timing, cookies, it measures and tells you if this JavaScript weights too much and it tells you to optimize it and that’s why we use Less. We have been using Less a lot recently.}---P06
\end{displayquote}

\begin{displayquote}
\say{jQuery is quite good for Sharepoint as we want to touch the server as [little] as possible and that’s why we put [a] bunch of code that creates the page without touching the server side and so the page is updated only by the engine of javascript. }---P17
\end{displayquote}

\highlight{A framework [documentation, author, website] should state clearly if it is intended for developing applications that will run relying mostly on the resources of the client or on those of the server.}\footnote{Several implications emerged from the data, and we first report them near the relevant evidence. }

\subsubsection*{Size} The size of an application developed via a JavaScript framework is used as a parameter for evaluating new frameworks to use: 

\begin{displayquote}
\say{I developed a small demo prototype to see how it [the framework] actually is practical, so how much code it produces for example.}---P12
\end{displayquote}

\highlight{The number of lines of code of the applications created by means of the framework should be as low as possible.}

\subsubsection{Effort expectancy}

Effort expectancy is the degree of ease associated with the use of the system~\cite{venkatesh2003}. Here we embrace the UTAUT definition, but we further characterize it by including those aspects that are related to when a framework reduces the effort and complexity when conducting programming tasks.\footnote{See footnote~\ref{explanation-performance}.}

\subsection*{Learnability} Reflects the effort that a developer should make to learn the framework. The requirements for starting to use a framework can be high or low but this also depends on the technical skills of the developer. Some participants stated that they choose the framework and libraries according to the time they are given to submit the project. 

\begin{displayquote}
\say{Most of the time you search for plugins to do something small and there you do not have to learn. You just check if it does what you need it to do. Otherwise, let's for instance refer to AngularJS it would take some weeks to learn the basics.}---P03
\end{displayquote}

\subsubsection*{Complexity} A perceived reduction of \textit{complexity} while developing is an important adoption factor.

\begin{displayquote}
Interviewer: \say{What do you mean by benefit? Is it in terms of a specific quality that they satisfy?}

P01: \say{We want to avoid complex systems and we want libraries [meant as frameworks] to integrate with what we already have.}
\end{displayquote}

\begin{displayquote}
Interviewer: \say{What did you like most of this framework?}

P08: \say{Every framework offers a new learning [experience,] the best feature is you don't need to be rockstar in the vanilla JS also these frameworks let you do many tasks which were earlier very very complex to implement which lead in saving the time, making the performance good and it [makes] the UI experience very good.}
\end{displayquote}

\subsubsection*{Understandability} Comprises the attributes of software that allow users to recognize how the JavaScript framework can be applied and used.

A well-documented framework enables time to be saved and more quickly extends the functionalities coming with the base source code.

\begin{displayquote}
\say{You also check the documentation if you want to do something which is not already available in the examples. Meaning, if you want to do something more extravagant.}---P03
\end{displayquote}

Additionally, developers look for running, documented examples to base their work on.

\begin{displayquote}
\say{There are a couple of methods that you should do an override and the rest it does automatically. There is a skeleton of the application and it guides you on how to implement the application. This is a big advantage according to me.}---P14
\end{displayquote}

\begin{displayquote}
\say{Most of these JavaScript frameworks come together with a set of examples and we [inspect] them to understand how we can use the package. If from the examples we notice that it seems easy enough then we take it into consideration otherwise we pick another one.}---P15
\end{displayquote}

\highlight{Documentation should be precise and include several examples for implementing common tasks. Documentation should allow developers to find guidelines on implementing a feature quickly.}

A framework with a clear code structure contributes to shortening the time a programmer needs to become familiar with the new platform.

\begin{displayquote}
\say{I got interested in AngularJS because of the two way data binding, and clean and easy to comprehend structure of the code. It is easy to become productive quite fast with this framework.}---P02 
\end{displayquote}

\highlight{The code that implements a framework should be easy to read and understand. At the same time, a framework should allow the creation of cleaner code as well.}

\begin{displayquote}
\say{I love the idea of directives. This makes it so much easier to reuse code, and the front-end code looks so much cleaner and easier to comprehend.}---P02
\end{displayquote}

\begin{displayquote}
\say{if you look at React you see a lot more computer science basics there, a lot of ideas that should help you structure your application, as well as use the most simplest approach like not mutating state and always having just one flow through your application and I think this is really beneficial.}---P10
\end{displayquote}

\highlight{A framework should enable the development of clear code.}

\subsubsection{Social influence}

Social influence is the degree to which an individual perceives that important other people either use a system, or believe that s/he should use the system~\cite{venkatesh2003}. Subjective norm, social influence and peer pressure, as well as competition~\cite{venkatesh2012} could drive the adoption of a technology.

\subsubsection*{Competitor analysis} Developers proxy the advice from competitor analyses, as follows:

\begin{displayquote}
 \say{We have also done some research on the market: how many companies, if there are other companies in similar context which are using Angular. And we found quite a lot of those.}---P12
\end{displayquote}

\highlight{A framework is estimated as reliable if used by similar companies in their production environment.}

\subsubsection*{Collegial advice}
The code represents those direct recommendations from peers and self-constructed networks of trust that practitioners employ when considering a framework.

Collegial advice is built by perceiving the popularity of a framework, as outlined by P04 and P18.

\begin{displayquote}
\say{People were always talking about the fantastic jQuery}---P04
\end{displayquote}

\begin{displayquote}
Interviewer: \say{Do you use the same principles [the proof of concept involves rebuilding websites with a new framework] when you have to choose libraries and frameworks for developing a specific functionality?}

P18: \say{Yes, the main principles involved, obviously we look at how popular is the library [referring to a framework] cause that helps indicate whether or not it has been used in existing projects.}
\end{displayquote}

Practitioners seek the Web for reviews, experiences, and opinions of peers, e.g.:

\begin{displayquote}
Interviewer: \say{So did you base your choice [jQuery and jQuery Mobile] also on blogs, on community or anything else?}

P07: \say{Particularly blogs, there were blogs which did comparisons on both the ease of use and the completeness of the arguments covered by the framework.}
\end{displayquote}

Practitioners suggested that, by looking at the presence of several mentions of the framework including examples and tutorial, they were more convinced by the choice.

\begin{displayquote}
\say{Before using any framework we study the framework by visiting the sites, searching video tutorials and once we are comfortable enough then only we use them.}---P08
\end{displayquote}

Developers also interact with colleagues, e.g.:

\begin{displayquote}
\say{If there is something particularly useful [\ldots] we can talk to the Ambassador of the area, in this case for example the Test Ambassador.}---P07
\end{displayquote}

\subsubsection*{Community size}
When talking about the attractiveness of JavaScript frameworks, the participants often mentioned the communities behind the development of the frameworks. 
Developers try to identify products that are expected to be used in the long term and that are improved continuously.
The community is the heart of an open-source project as it is a container for new ideas to extend product functionalities, and bug fixes. The larger the community, the faster the response time in issue resolution. We developed two codes related to the community, namely community size and (next) community responsiveness.

\begin{displayquote}
Interviewer: \say{Which parameters did you check or did you test for your study [referring to a search for a JavaScript framework]?}

P12: \say{Well it was mainly the community size was one criteria, then the development tools which you had available for the framework.}

Interviewer: \say{How come you changed from Java MVC to Angular JS?}

P12: \say{Well, basically it is for a new product we are doing and there we are planning to potentially extract some source code parts of the product into the open source community and therefore Angular JS is much more convenient. Because the community around it is much bigger. }

Interviewer: \say{So you mean the people who are contributing?}

P12: \say{Yes, the people who are contributing. And then it is more modern let’s say [\ldots]}
\end{displayquote}

\begin{displayquote}
Interviewer: \say{Currently which framework would you advise more [...]?}

P12: \say{Currently the focus is for sure on Angular but also Ember is quite diffused. The problem is with Ember it has for example a quite young community so you do not really know yet how it is going in the long term.}
\end{displayquote}

\begin{displayquote}
P10: \say{The React ecosystem is not as large as the Ember or Angular ecosystem it still needs some work on a few thing[s] like having components for tag input or day pickers but nothing really major that everybody is using, so I think React especially could benefit from a larger community and have more influence there.}
\end{displayquote}

\subsubsection*{Community responsiveness}
As mentioned above, community size went hand-in-hand with the perceived responsiveness of the community. Developers look forward to active communities, because it is expected that issues are solved quickly.

\begin{displayquote}
\say{Communities are quite important, we rely on them every time and, [the bigger] the community [, the] faster you receive responses from them. [\ldots] When querying a community of open source for such big projects such as AngularJS it is easy that you receive a fast response very easily [and quickly].}---P01
\end{displayquote}

\highlight{The community behind a framework is its heart and as such its size and contributors signal the trust of developers in the technology.}

Decision makers place value on the age of a framework as a proxy of the work of the community behind it.

\begin{displayquote}
Interviewer:\say{Do you see Angular as a mature enough framework?}

P12: \say{Yes, I would say yes. }

Interviewer: \say{Because it has some characteristics like you said: the community and what else comes in your mind?}

P12: \say{Also because of its age probably because it is the oldest framework which is around. It took years to get popular and now in these years it took off. So that’s probably the [most important] reason and then also compared to the features. So it has already some advanced features in respect with the other frameworks. }

\say{I think AngularJS is a mature framework also because of its age probably because it is the oldest framework which is around. It took years to get popular and now in these years it took off.}---P12 
\end{displayquote}

\begin{displayquote}
\say{Every framework with time gets mature when many people around the world try their hands in that.}---P08
\end{displayquote}

\highlight{Frameworks which have been around for a longer time are preferred to new ones.}

\subsubsection{Facilitating conditions}
Facilitating conditions are defined as the degree to which an individual believes that an organizational and technical infrastructure exists to support the use of the system~\cite{venkatesh2003}. In our specific case, it collects those framework characteristics in the context of suiting current requirements and integration potential.

\subsubsection*{Suitability} Collects the attributes which express if the product is appropriate for fulfilling the tasks required in a specific function.

One of the front-end programmers stated that searching for a framework or a functionality is mostly looking for what problem it is solving. Several others pointed out that nowadays libraries are commonly included in the native libraries of a framework.

\begin{displayquote}
\say{If I go for Knockout JavaScript if I want to go for HTTP connectivity I will check if jQuery has something but when you take AngularJS it has its own library to do that.}---P13
\end{displayquote}

\highlight{Simple tasks such as event handling, DOM manipulation, and real time component updates should be automatized.}

A framework which can create a complete application fulfilling the expectations of the developer gains advantage over others; it is seen as a good solution for saving time and avoiding exhaustive search for the suitable functionality. 

\begin{displayquote}
\say{[referring to Dojo Toolkit] I chose this framework because I needed something which could create complete applications and something that could handle the chaos that was by that time of the JavaScript libraries and this framework does both of these things in a decent way.}---P11 
\end{displayquote}

Decision makers evaluate JavaScript frameworks over time and according to tasks and needs. They can make proposals to their team leader (or the rest of the team in case of self-organized teams). The team leader (the team) analyzes the request to identify how it could possibly be integrated into the project. If the integration is feasible, the leader (team) accepts the request or identifies libraries which are already provided, which can achieve the desired functionalities.

\begin{displayquote}
P07: \say{If there is something particularly useful [... and] if it implements something interesting they [the frameworks] are inserted in the project. Then they [team leaders] organize tech talks to teach how to use the new added libraries [meant as framework functionalities].} 

Interviewer: \say{During the tech talks do you discuss the structure of a library, functionalities or what else?}

P07: \say{Yes, for instance I needed to use PowerMojito which is an extension of Mojito. I would go to the ambassador which takes care of the test section. If I go there and tell him that using it has \textit{this and that} advantages, he can decide to insert it in the libraries [meant as framework functionalities] which are imported by default. He then decides to arrange tech talks, during which they talk about the functionalities and why it is useful and how to use it.}
\end{displayquote}

The framework should not only be suitable to achieve a requirement, it should also be suitable for the software engineers. A developer stated that they also appreciate the architecture of the framework as it allows him to have continuous control and full visibility of the structure, and this is one of his needs. 

\begin{displayquote}
\say{I am more and more moving toward React because it is solving the problems in a more elegant way, in my opinion. So organizing the rendering strategy between all your components and parts of the application is way more organized if you use React.}---P10
\end{displayquote}

\subsubsection*{Updates} Describes the need of developers to use frameworks which (1) are continuously updated and (2) extend their functionalities to stay competitive in the market.

\begin{displayquote}
\say{And then AngularJS is more modern let's say. So it has some features which Java MVC didn't have or doesn't yet have.}---P12
\end{displayquote}

\begin{displayquote}
\say{[\ldots] also [frameworks should] allow the implementation of new technologies}---P04
\end{displayquote}

\highlight{Frequently updating a framework with new features for matching web design trends is evaluated positively.}

Updates ensure compatibility, as well.

\begin{displayquote}
\say{The issue was that MooTools did not have a continuous evolution and had a lot of bugs and issues with different browsers.}---P06
\end{displayquote}

\highlight{Applications created through a framework should run in different browsers.}

\subsubsection*{Modularity} One of the developers who was not using any framework for developing but was integrating libraries into pure JavaScript code, stated that the principle of modularity does not apply when using your own code as sometimes issues of compatibility can occur:

\begin{displayquote}
Interviewer: \say{Did it ever happen that they did not fully integrate with what you were developing?}

P09: \say{You bet. This is not so rare. Sometime, I have to look around and implement another one library or to hack and modify the source code.}
\end{displayquote}

A developer who had built their own framework pointed out that another crucial aspect was the need for extensibility and modularity in the sense of scalability, that was the reason they programmed a new framework for JavaScript. 

\begin{displayquote}
\say{The framework is composed of different libraries, allowing scalability and also allows users to use their own libraries [meant as framework component], making them part of the core system [\ldots] It allows to change one module without affecting the others.}---P04
\end{displayquote}

\highlight{A framework should be modular; changing a module should not affect others.}

The modularity of a framework means that it is flexible, too. 
It should be easy to import libraries and features, as portrayed by P12.

\begin{displayquote}
\say{As far as I know there is no really good approach for modularity in the application. So if you have, if you want to implement some more advanced features like lazy loading of modules and that stuff you have to do it on your own. [\ldots] AngularJS allows importing external libraries because most of the parts are based on jQuery, however you have to wrap those into Angular directives. So that's for sure one disadvantage of Angular. [\ldots] you cannot simply take a jQuery plugin for instance and put it into the project. You could, but it is not suggested by the Angular team. You have to wrap it such that you can augment those plugins by augmenting the initial HTML code basically. So through those Angular directives. That's how they call it.}---P12
\end{displayquote}

Another type of flexibility achieved trough modularity is the possibility to add a framework at a later stage of development.

\begin{displayquote}
\say{I start with pure JS to do the basic and necessary stuff and then use framework or lib.}---P09
\end{displayquote}

\highlight{Libraries that come with a framework should enable the achievement of basic and advanced functionalities at any development stage.}

\subsubsection*{Isolation} The participants declared that they tend to develop applications that exploit the client as much as possible instead of the server. This design strategy is done to preserve the server status.

\begin{displayquote}
\say{We want to touch the server as little as possible and to keep the application mostly on the client side, that's why we put [a] bunch of code that creates the page without touching the server side and so the page is updated only by the engine of JavaScript.}---P17
\end{displayquote}

\subsubsection*{Extensibility} Pointed out by a participant when talking about issues in extending frameworks: 

\begin{displayquote}
\say{Let's say it is easier to use [Angular] but more difficult to extend. Let's say there are other frameworks which are more difficult to use but then are more extensible.}---P03
\end{displayquote}

\begin{displayquote}
\say{If something is not already there in Angular JS it is flexible enough to use external libraries as well.}---P13
\end{displayquote}

\highlight{A framework should allow external libraries to be imported without having to adapt them.}

\subsubsection{Price value}

Price value was not included in the original UTAUT, yet our participants often mentioned the costs of adopting a framework. Indeed, price value was included in a revision of UTAUT~\cite{venkatesh2012} and we adopted it as one of our major categories.

\subsubsection*{Cost}

Five of the participants stated that they try to use mostly free frameworks. The participants stated that this depends on the customer's requirement to keep the costs low: 

\begin{displayquote}
\say{If something [referring to frameworks and libraries] is not free then we see how much it costs, if it is bearable for the company and plus what functionality it is providing to us.}---P13
\end{displayquote}

\begin{displayquote}
\say{The libraries we use are chosen first of all [from] the open source [community]. The customers prefer projects which are composed of many open source components.}---P15
\end{displayquote}

The developers motivated the choice either as coming from a company's strategy or as coming from the customer's requirements. 

\begin{displayquote}
\say{I proposed the change due to the fact that we have to go open source}---P12
\end{displayquote}

\highlight{Free frameworks are preferred to paid ones.}

\subsection{Actors leading to the adoption of a JavaScript Framework}
\label{ssec:results:actors}
\begin{figure*}
\centering
\includegraphics []{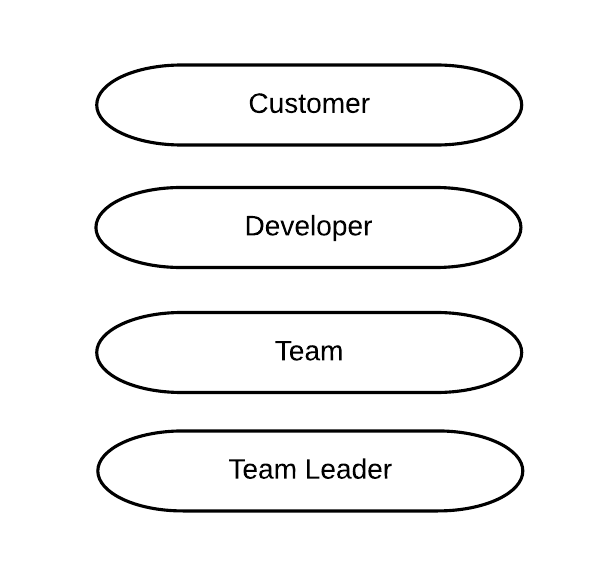}
\caption{Actors leading to the adoption of a JavaScript Framework}\label{fig:model-actors}
\end{figure*}

The codes for the \textit{Actors} central category refer to how the developers decide which framework to proceed with or which library to include during the development. The categories that we identified are displayed in  Figure~\ref{fig:model-actors}.

Developers affirmed that the solution is based on the customer's decision in case they had to work on existing projects or chosen freely according to the parameters presented in the previous section. It was interesting for the purpose of the research to explore the adoption of frameworks in teams. Two categories arose: decisions made as a team and decisions made by the team leader. 

Teams adopt different strategies for evaluating frameworks, as stated in the following interview snippet.

\begin{displayquote}
\say{We took an existing set of pages and rebuilt them using Angular and what we looked at was the amount of effort involved and how easily testable it was. We check in team the libraries [meant as frameworks] which could work good for solving an issue and we meet and talk, as the community is quite big and it evolves quickly. We choose the one library which could bring us more benefit. Nothing is rushed because there is too much out there.}---P18 
\end{displayquote}

\begin{displayquote}
\say{We have a team that researches and then we propose the solutions to the others so that each one can say his opinion. We state the characteristics of each and then we discuss which one to pick.}---P01 
\end{displayquote}

\begin{displayquote}
\say{If there is something particularly useful which is not being covered by the libraries [meant as frameworks], which have already been imported, we can talk to the Ambassador of the area, in this case for example the Test Ambassador. You show him the plus and minus. If you need something else, for instance recently, I needed the framework we use for doing tests. The one we use to create models of computations.}---P07
\end{displayquote}

\section{Discussion}
\label{sec:discussion}

In this section, we first compare our results to the related work in the scientific literature and the practitioner viewpoints. Then, we state the theoretical and practical implications of the results. Finally, we elaborate on the limitations of the study.
 
\subsection{Scientific literature}

Our findings confirm that the size of a framework, as suggested by Misra and Cafer~\cite{Misra2012e} and Gizas et. al~\cite{Gizas2012d}, is important for practitioners, but it is not sufficient alone as a metric. Practitioners care to understand the payload brought by frameworks in the context of their own web applications as well. For this purpose, developers implement ad-hoc benchmarks to measure the number of lines of code generated for a sample web page or a sample application. The performance of the framework measured in terms of execution time, as introduced by Gizas et. al~\cite{Gizas2012d} and Misra and Cafer~\cite{Misra2012e}, is confirmed to be a factor influencing the choice of a JavaScript framework. 

This research corroborates the statement by Graziotin and Abrahamsson~\cite{Graziotin2013bd} that decision makers study the documentation to evaluate the time they would need to be productive. More specifically, developers tend to prefer frameworks that include examples of achieving simple tasks and hints for achieving advanced functionalities. Additionally, our research supports the classification of a frameworks' maturity introduced by Graziotin and Abrahamsson~\cite{Graziotin2013bd}, based upon the frequency of released updates. This classification is extended by adding the frameworks' age parameter. This parameter was also mentioned as an important factor by the technical blogger Reyes~\cite{Reyes2010b}. 

Stefi~\cite{stefi2015} employed UTAUT for constructing a quantitative survey investigating the intention to adopt software components. While a direct comparison with our work is a difficult task, we can report that our results support the findings that UTAUT performance expectancy and social influence have a role in the adoption decision of JavaScript frameworks as well.

The work of Polan\v{c}i\v{c} et al.~\cite{polancic2010} has a very similar context to our work; however, they employed a different theory for constructing their instrument (TAM instead of UTAUT). We attempt to compare results here, too, given that UTAUT was born as a TAM successor. The most impressive results of the study showed a strong relationship between the perceived usefulness and ease of use of a framework and its adoption. We found supporting evidence in the qualitative data, given that several of our codes were in the similar categories of facilitating conditions, effort expectancy and performance expectancy. Also, Polan\v{c}i\v{c} et al. concluded that ``framework developers should be aware that the acceptance of frameworks is dependent on framework users perceptions.'' (p. 582), which we fully support in our study. Developers are users of frameworks, and produce for users or frameworks. This is why it is important to understand which characteristics of a JavaScript framework drive its adoption, for producing future frameworks.

\subsection{Practitioner viewpoints}

A JavaScript framework, as mentioned in the technical blog of Bennet~\cite{Bennett2007c}, should be accurate and represent the language as it is. This, according to our findings, can be achieved if the framework allows developers to have an overview of the execution flow of applications. Developers like to understand how the framework operates in order to promptly identify possible issues. 

Our results confirm that modularity, as Bennet~\cite{Bennett2007c} reported, is a factor that plays a significant role in the choice of the framework. Modular frameworks ease the process of the modification of code sections without affecting other components. Our findings match the statement of Walsh~\cite{Walsh2007c} that a framework should be suitable for achieving the functionality it addresses. Furthermore, decision makers prefer frameworks which include libraries for achieving basic and advanced functionalities in their core version. 

Our research outlines the factors which influence the choice of a JavaScript framework, namely performance expectancy (performance, size), effort expectancy (automatization, learnability, complexity, understandability), social influence (competitor analysis, collegial advice, community size, community responsiveness), facilitating conditions (suitability, updates, modularity, isolation, extensibility), and price value. These factors are evaluated by a combination of four possible decision makers: customer; developer; team; and team leader.

\subsection{Implications}
\label{sec:implications}

Our research enriches the body of knowledge of technology adoption of software engineers. Our taxonomy is organized with UTAUT but, on the lower level, identifies 16 attributes that are relevant to the choice of a JavaScript framework. Several of these factors are novel, such as the native support for basic and advanced features, the compatibility with existing infrastructures, the possibility to choose if the application should run mostly on the client or on the server side, and its cost. To our knowledge, this is the first qualitative study in the domain. As such, it sets basic building blocks for quantitative and qualitative research on the factors we identified. Our model can also be exploited by future studies on the process of adopting a JavaScript framework. While our research has focused on JavaScript frameworks exclusively, future studies can verify the suitability of our model regarding the adoption of frameworks for other programming languages.

The results of our study favor practical implications. Decision makers can employ the model that we propose as a reference to evaluate new JavaScript frameworks or to confirm the choice of the already chosen ones. We are not providing any quantitative approach to drive the selection, as this is left to future studies. Practitioners are free to select how to measure the adherence of the candidate framework to the factors we discovered, thus guiding the decision.

Our model can also be inspiring for those developers willing to publish new frameworks so as to make them captivating. The variables introduced by the model reflect the thoughts and expectations of practitioners which have a broad view on the portfolio of JavaScript frameworks in the market. The development of a new framework could thus try to observe these recommendations so as to gain popularity with respect to already existing ones. 
 
Finally, this research has outlined a number of interesting observations that we offer in Table~\ref{tbl:implications}. The implications are mostly directed at developers for the creation{} of desirable frameworks. However, we believe that they can also be read by decision makers as guidelines for selecting a framework, together with the taxonomy we produced.

\begin{boxed}
\begin{todolist}
\item A framework [documentation, author, website] should state clearly if it is intended for developing applications that will run relying mostly on the resources of the client or on those of the server.
\item The number of lines of code of the applications created by means of the framework should be as low as possible.
\item Documentation should be precise and include several examples for implementing common tasks. Documentation should allow developers to find guidelines on implementing a feature quickly.
\item The code that implements a framework should be easy to read and understand.
\item A framework should enable the development of clear code.
\item A framework is estimated as reliable if used by similar companies in their production environment.
\item The community behind a framework is its heart and as such its size and contributors signal the trust of developers in the technology.
\item Frameworks which have been around for a longer time are preferred to new ones. 
\item Simple tasks such as event handling, DOM manipulation, and real time component updates should be automatized.
\item Frequently updating a framework with new features for matching the web design trends is evaluated positively.
\item Applications created through a framework should run in different browsers. 
\item A framework should be modular; changing a module should not affect others.
\item Libraries that come with a framework should enable the achievement of basic and advanced functionalities at any development stage.
\item A framework should allow external libraries to be imported without having to adapt them.
\item Free frameworks are preferred to paid ones.
\end{todolist}
\captionof{table}{Implications for JavaScript Frameworks developers and decision makers\label{tbl:implications}}
\end{boxed}

It is worth noting that the previously stated implications have a stronger meaning when singled out. 
When considered on the whole, the implications might carry cross-tradeoffs. 
For example, a framework with an easy to understand code could be a paid one.
Analyzing the tradeoffs is outside the scope of this study. 
We leave our remarks as a suggestion for future research.

\subsection{Limitations}

The most significant limitation of this research is that the sample is limited. Although we did sample up to an agreement that there was theoretical saturation, we decided to sample only for expert individuals, not newcomers. Our 18 participants were recruited using ever refining criteria, and we were able to gather a rich amount of data. We were able to reach theoretical saturation~\cite{Strauss1994b}. Additionally, the sample size belongs to the range of the suggested number of participants in qualitative studies~\cite{Guest2006c}.

Another limitation of this study is that we relied on our participants' own judgment for what they considered a JavaScript framework. In particular, some participants declared to use in-house frameworks without disclosing much about them.

A minor limitation of this study is that the interviews were done mostly using Skype, Google Hangouts or on the phone. This kind of communication differs from the face-to-face interview. The environment typically chosen for the interviews done in person would also have allowed more focus on the emotions and expressions of the interviewees. Conversely, the participants are all familiar with these instant messaging and conferencing technologies and feel more comfortable if they choose the time and the location for the interview. Besides this, as the participants who were recruited work in different countries it would have been economically unfeasible to arrange interviews with all of them in person. 

\section{Conclusions}
\label{sec:conclusions}

The proliferation and diversity of JavaScript applications has led to a plethora of JavaScript frameworks. Novices and experienced programmers face several challenges to identify the patterns they should take into consideration when choosing a framework. Furthermore, developers who wish to release new frameworks require an overview of the most relevant characteristics that a framework must ensure to guarantee its adoption. 

The present work identified a lack of research for solving the previously reported issues when selecting a JavaScript framework. Existing research has mainly focused on predefined metrics which have been reported to have little meaning for practitioners. While benchmarks, technical reports, and expert's opinions are available, they suffer the same problem identified in the scientific research field. The knowledge we were missing concerned the factors that drive developers towards the adoption. 

The purpose of our study was to identify and understand the factors that influence the choice of a JavaScript framework with respect to another. In this paper, we reported a qualitative interpretive study of semi-structured interviews. We interviewed eighteen participants who are decision makers in their companies or on their own, or are able to motivate the JavaScript framework decision and adoption. 

Our research aimed to answer the research question \textit{Which factors and actors lead to the adoption of a JavaScript framework?} 

Through a qualitative coding of the interview responses, we offer a model of desirable JavaScript framework adoption factors. The factors are grouped into categories that are derived by the Unified Theory of Acceptance and Use of Technology (UTAUT). The factors are performance expectancy (performance, size), effort expectancy (automatization, learnability, complexity, understandability), social influence (competitor analysis, collegial advice, community size, community responsiveness), facilitating conditions (suitability, updates, modularity, isolation, extensibility), and price value. A combination of four actors, who are customer, developer, team, and team leader, leads to the choice.

Our work confirmed those technical factors which had already been discovered in previous research. However, we found that the factors introduced by prior research were only the tip of the iceberg for understanding the adoption and are often assessed in ways that are not meaningful for practitioners.

The present paper introduces several implications for framework developers and for the decision makers in the context of technology adoption. Our classification offers desirable factors (Figure~\ref{fig:model-factors}) and suggestions (Section~\ref{sec:implications}) that can drive the development of future JavaScript frameworks. On the other hand, the same factors and suggestion can help current decision makers in the adoption of existing JavaScript frameworks.

Our work lays down foundations for future research. Quantitative studies can be conducted in order to employ or develop metrics for each of the discovered attributes we reported in Section~\ref{sec:results:factors} and Section~\ref{ssec:results:actors}. 
Future works should lead towards the creation of benchmarks that have real, practical utility. Altogether, we envision the creation of a global ranking system of JavaScript frameworks which will be valued by the practitioners. We believe, however, that such a system should be created only after the development of metrics that operationalize our factors.

\section{Acknowledgments}
We would like to thank all the participants of this study.

Our gratitude goes to the editor and three anonymous reviewers whose feedback improved our original
efforts immeasurably.

Daniel Graziotin has been supported by the Alexander von Humboldt (AvH) Foundation.

\bibliographystyle{spmpsci}
\bibliography{Bibliography} 

\begin{thebibliography}{10}
\providecommand{\url}[1]{{#1}}
\providecommand{\urlprefix}{URL }
\expandafter\ifx\csname urlstyle\endcsname\relax
  \providecommand{\doi}[1]{DOI~\discretionary{}{}{}#1}\else
  \providecommand{\doi}{DOI~\discretionary{}{}{}\begingroup
  \urlstyle{rm}\Url}\fi

\bibitem{vishwanath2016a}
Understanding How Social-Behavioural Science Theory Can Explain the Design of
  Software Websites, vol. 2016 49th Hawaii International Conference on System
  Sciences (HICSS). IEEE (2016)

\bibitem{angularjs}
AngularJS: Angularjs (2017).
\newblock \urlprefix\url{https://angularjs.org}

\bibitem{W3CSS}
Atkins, T.J., Etemad, E.J., Rivoal, F.: Css snapshot 2015.
\newblock Tech. rep., W3C (2015).
\newblock \urlprefix\url{https://www.w3.org/TR/CSS/}

\bibitem{Bennett2007c}
Bennett, J.: {Choosing a JavaScript library} (2007).
\newblock
  \urlprefix\url{http://www.b-list.org/weblog/2007/jan/22/choosing-javascript-library/}

\bibitem{meyer2014}
Cheung, S.C., Orso, A., Storey, M.A. (eds.): Software developers' perceptions
  of productivity, vol. the 22nd ACM SIGSOFT International Symposium. ACM
  Press, New York, New York, USA (2014)

\bibitem{Craver2017}
Craver, N.: New york is so cool for programmers, they even have this sign that
  keeps track of how many javascript frameworks are out there (2017).
\newblock
  \urlprefix\url{https://web.archive.org/web/20170717154926/https:/twitter.com/Nick_Craver/status/886192620065705984}

\bibitem{Creswell2009}
Creswell, J.W.: {Research design: qualitative, quantitative, and mixed method
  approaches}, vol. 2nd, 3 edn.
\newblock Sage Publications, Thousand Oaks, California (2009)

\bibitem{Cunha:2016kw}
Cunha, J., Moura, H.P., Vasconcellos, F.: {Decision-making in Software Project
  Management: A Systematic Literature Review}.
\newblock Procedia Computer Science \textbf{100}, 947--954 (2016)

\bibitem{r02166}
Davis, F.D., Bagozzi, R.P., Warshaw, P.R.: User acceptance of computer
  technology: a comparison of two theoretical models.
\newblock Management science \textbf{35}(8), 982–1003 (1989)

\bibitem{designzum}
designzum.com: designzum.com (2017).
\newblock \urlprefix\url{http://designzum.com}

\bibitem{dojo}
Dojo: Dojo (2017).
\newblock \urlprefix\url{https://dojotoolkit.org}

\bibitem{doukakis2013}
Doukakis, S., Giannakos, M.N., Koilias, C., Vlamos, P.: Measuring students’
  acceptance and confidence on algorithms and programming: The impact of the
  engagement with cs on secondary education.
\newblock Informatics in Education \textbf{12}(2), 207–219 (2013)

\bibitem{Dreyfus1987b}
Dreyfus, H.L., Dreyfus, S.E., Zadeh, L.A.: {Mind over Machine: The Power of
  Human Intuition and Expertise in the Era of the Computer}.
\newblock IEEE Expert \textbf{2} (1987).
\newblock \doi{10.1109/MEX.1987.4307079}

\bibitem{easterbrook2008}
Easterbrook, S., Singer, J., Storey, M.a., Damian, D.: {Selecting empirical
  methods for software engineering research}.
\newblock In: Guide to Advanced Empirical Software Engineering, pp. 285--311.
  Springer (2008).
\newblock \doi{10.1007/978-1-84800-044-5\_11}

\bibitem{extjs}
ExtJS: Extjs (2017).
\newblock \urlprefix\url{https://www.sencha.com/products/extjs/}

\bibitem{facebook}
Facebook: Facebook (2017).
\newblock \urlprefix\url{https://facebook.com}

\bibitem{fagerholm2015}
Fagerholm, F., Ikonen, M., Kettunen, P., M{\"u}nch, J., Roto, V., Abrahamsson,
  P.: Performance alignment work: How software developers experience the
  continuous adaptation of team performance in lean and agile environments.
\newblock Information and Software Technology \textbf{64}, 132--147 (2015)

\bibitem{Fayad1997c}
Fayad, M., Schmidt, D.C.: Object-oriented application frameworks.
\newblock Commun. ACM \textbf{40}(10), 32--38 (1997).
\newblock \doi{10.1145/262793.262798}

\bibitem{garrett2005ajax}
Garrett, J.J.: Ajax: A new approach to web applications  (2005).
\newblock Available at http://archive.is/NodKD

\bibitem{geertz1973}
Geertz, C.: {The Interpretation of Cultures: Selected Essays}, vol.~1.
\newblock Basic Books, New York, New York, USA (1973)

\bibitem{Gizas2012d}
Gizas, A.B., Christodoulou, S.P., Papatheodorou, T.S.: {Comparative evaluation
  of javascript frameworks}.
\newblock In: Proceedings of the 21st international conference companion on
  World Wide Web, pp. 513--514 (2012)

\bibitem{glaser2014}
Glaser, B.G.: Choosing grounded theory.
\newblock Grounded Theory Review \textbf{13}(2), 3–19 (2014)

\bibitem{googleplus}
Google: Google plus (2017).
\newblock \urlprefix\url{https://plus.google.com}

\bibitem{Graziotin2013bd}
Graziotin, D., Abrahamsson, P.: Making sense out of a jungle of javascript
  frameworks - towards a practitioner-friendly comparative analysis.
\newblock In: Product-Focused Software Process Improvement - 14th International
  Conference, {PROFES} 2013, Paphos, Cyprus, June 12-14, 2013. Proceedings, pp.
  334--337 (2013).
\newblock \doi{10.1007/978-3-642-39259-7\_28}

\bibitem{Guest2006c}
Guest, G.: How many interviews are enough?: An experiment with data saturation
  and variability.
\newblock Field Methods \textbf{18}(1), 59--82 (2006).
\newblock \doi{10.1177/1525822X05279903}

\bibitem{Halstead1977c}
Halstead, M.: {Elements of Software Science (Operating and programming systems
  series)}.
\newblock Elsevier Science Inc., New York, NY, USA (1977)

\bibitem{hardgrave2003a}
Hardgrave, B., Johnson, R.: Toward an information systems development
  acceptance model: The case of object-oriented systems development.
\newblock IEEE Trans. Eng. Manage. \textbf{50}(3), 322–336 (2003)

\bibitem{heath2004}
Heath, H., Cowley, S.: Developing a grounded theory approach: a comparison of
  glaser and strauss.
\newblock International Journal of Nursing Studies \textbf{41}(2), 141–150
  (2004)

\bibitem{hong2011a}
Hong, W., Thong, J.Y.L., Chasalow, L.C., Dhillon, G.: User acceptance of agile
  information systems: A model and empirical test.
\newblock Journal of Management Information Systems \textbf{28}(1), 235–272
  (2011)

\bibitem{github}
https://github.com: Github (2017).
\newblock \urlprefix\url{https://github.com}

\bibitem{githubstarredrepos}
https://github.com: Most starred github repositories (2017).
\newblock
  \urlprefix\url{https://github.com/search?q=stars%3A%3E1&type=Repositories}

\bibitem{idemudia2016a}
Idemudia, E.C., Dasuki, S.I., Ogedebe, P.: Factors that influence students’
  programming skills: a case study from a nigerian university.
\newblock IJQRE \textbf{3}(4), 277 (2016)

\bibitem{nvivo}
International, Q.: Nvivo 10 (2012).
\newblock \urlprefix\url{http://www.qsrinternational.com/nvivo-product}

\bibitem{jquery}
jQuery: jquery (2017).
\newblock \urlprefix\url{https://jQuery.com}

\bibitem{jsdb}
jsdb.io: jsdb.io (2017).
\newblock \urlprefix\url{http://jsdb.io}

\bibitem{jster}
Jster.net: Jster.net (2017).
\newblock \urlprefix\url{http://jster.net}

\bibitem{W3XMLHttpRequest}
van Kesteren, A., Aubourg, J., Song, J., Steen, H.R.M.: Xmlhttprequest level 1.
\newblock Tech. rep., W3C (2014).
\newblock
  \urlprefix\url{https://www.w3.org/TR/2014/WD-XMLHttpRequest-20140130/}

\bibitem{W3DOM}
van Kesteren, A., Gregor, A., Ms2ger, Russell, A., Berjon, R.: W3c dom4.
\newblock Tech. rep., W3C (2015).
\newblock \urlprefix\url{http://www.w3.org/TR/2015/REC-dom-20151119/}

\bibitem{kettunen2013}
Kettunen, P.: The many facets of high-performing software teams: A
  capability-based analysis approach.
\newblock Systems, Software and Services Process Improvement pp. 131--142
  (2013)

\bibitem{klein1999}
Klein, H.K., Myers, M.D.: {A Set of Principles for Conducting and Evaluating
  Interpretive Field Studies in Information Systems}.
\newblock MIS Quarterly - Special issue on intensive research in information
  systems \textbf{23}(1), 67 (1999).
\newblock \doi{10.2307/249410}

\bibitem{Langley1999}
Langley, A.: {Strategies for Theorizing from Process Data}.
\newblock The Academy of Management Review \textbf{24}(4), 691 (1999).
\newblock \doi{10.2307/259349}

\bibitem{linkedin}
LinkedIn: Linkedin (2017).
\newblock \urlprefix\url{https://linkedin.com}

\bibitem{mccabe1976complexity}
McCabe, T.J.: A complexity measure.
\newblock IEEE Transactions on software Engineering (4), 308--320 (1976)

\bibitem{McNamara1999b}
McNamara, C.: {General Guidelines for Conducting Interviews. In Field Guide to
  Consulting and Organizational Development}.
\newblock Authenticity Consulting, LLC  (1999).
\newblock
  \urlprefix\url{http://managementhelp.org/businessresearch/interviews.htm}

\bibitem{Misra2012e}
Misra, S., Cafer, F.: {Estimating Quality of JavaScript}.
\newblock The International Arab Journal of Information Technology
  \textbf{9}(November), 535--543 (2012)

\bibitem{Mohr1982}
Mohr, L.B.: {Explaining Organizational Behavior}.
\newblock Jossey-Bass Inc (1982)

\bibitem{mootools}
MooTools: Mootools (2017).
\newblock \urlprefix\url{http://mootools.net/}

\bibitem{nodejs}
NodeJS: Nodejs (2017).
\newblock \urlprefix\url{https://nodejs.org}

\bibitem{Ocariza2013b}
Ocariza, F., Bajaj, K., Pattabiraman, K., Mesbah, A.: {An Empirical Study of
  Client-Side JavaScript Bugs}.
\newblock 2013 ACM / IEEE International Symposium on Empirical Software
  Engineering and Measurement pp. 55--64 (2013).
\newblock \doi{10.1109/ESEM.2013.18}

\bibitem{gasparic2017}
Papadopoulos, G.A., Kuflik, T., Chen, F., Duarte, C., Fu, W.T. (eds.): GUI
  Design for IDE Command Recommendations, vol. the 22nd International
  Conference. ACM Press, New York, New York, USA (2017)

\bibitem{Pfeffer1983}
Pfeffer, J.: {Reviewed Work: Explaining Organizational Behavior. by Lawrence B.
  Mohr.}
\newblock Administrative Science Quarterly \textbf{28}(2), 321 (1983).
\newblock \doi{10.2307/2392635}.
\newblock \urlprefix\url{http://www.jstor.org/stable/2392635?origin=crossref}

\bibitem{polancic2010}
Polančič, G., Heričko, M., Rozman, I.: An empirical examination of
  application frameworks success based on technology acceptance model.
\newblock Journal of Systems and Software \textbf{83}(4), 574–584 (2010)

\bibitem{prototypejs}
Prototype: Prototype (2017).
\newblock \urlprefix\url{http://prototypejs.org}

\bibitem{Reyes2010b}
Reyes, J.: {How to Choose a Right JavaScript Framework} (2010).
\newblock
  \urlprefix\url{http://designreviver.com/tips/how-to-choose-a-right-javascript-framework/}

\bibitem{Robins2003c}
Robins, A., Rountree, J., Rountree, N.: {Learning and Teaching Programming : A
  Review and Discussion}.
\newblock Computer Science Education \textbf{13}, 137--172 (2003).
\newblock \doi{10.1076/csed.13.2.137.14200}

\bibitem{stefi2015}
Stefi, A.: Do developers make unbiased decisions? - the effect of mindfulness
  and not-invented-here bias on the adoption of software components.
\newblock ECIS 2015 Completed Research Papers. p. 175 (2015)

\bibitem{stol2016}
Stol, K.J., Ralph, P., Fitzgerald, B.: Grounded theory in software engineering
  research.
\newblock the 38th International Conference p. 120–131 (2016)

\bibitem{Strauss1994b}
Strauss, A., Corbin, J.: Grounded theory methodology.
\newblock Handbook of qualitative research \textbf{17}, 273--285 (1994)

\bibitem{strauss1998}
Strauss, A., Corbin, J.: Basics of Qualitative Research: Techniques and
  Procedures for Developing Grounded Theory.
\newblock SAGE Publications, Inc (1998)

\bibitem{Strauss2008b}
Strauss, A., Corbin, J.: {Basics of Qualitative Research (3rd ed.): Techniques
  and Procedures for Developing Grounded Theory}, vol.~3.
\newblock SAGE Publications, Inc., 2455 Teller Road, Thousand Oaks California
  91320 United States (2008).
\newblock \doi{10.4135/9781452230153}

\bibitem{mootoolsspeed}
test, M.S.S.: Mootools slick speed test (2017).
\newblock \urlprefix\url{https://github.com/mootools/slick/tree/master/speed}

\bibitem{Tiobeb}
Tiobe: {Tiobe Index} (2015).
\newblock
  \urlprefix\url{http://www.tiobe.com/index.php/content/paperinfo/tpci/index.html}

\bibitem{twitter}
Twitter: Twitter (2017).
\newblock \urlprefix\url{https://twitter.com}

\bibitem{VanDeVen1995}
{Van De Ven}, A.H., Poole, M.S.: {Explaining Development and Change in
  Organizations}.
\newblock Academy of Management Review \textbf{20}(3), 510--540 (1995)

\bibitem{venkatesh2003}
Venkatesh, V., Morris, M.G., Davis, G.B., Davis, F.D.: User acceptance of
  information technology: Toward a unified view.
\newblock MIS Quarterly \textbf{27}(3), 425–478 (2003)

\bibitem{venkatesh2012}
Venkatesh, V., Thong, J.Y.L., Xu, X.: Consumer acceptance and use of
  information technology: Extending the unified theory of acceptance and use of
  technology.
\newblock MIS Q. \textbf{36}(1), 157–178 (2012)

\bibitem{W3Techsd}
W3Techs: {Web Technology Surveys - Usage of JavaScript for websites} (2014).
\newblock
  \urlprefix\url{http://w3techs.com/technologies/details/cp-javascript/all/all}

\bibitem{Walsh2007c}
Walsh, D.: {8 Considerations For Choosing Your JavaScript Framework} (2007).
\newblock
  \urlprefix\url{http://davidwalsh.name/javascript-framework-considerations}

\bibitem{Walsham2006}
Walsham, G.: {Doing interpretive research}.
\newblock European Journal of Information Systems pp. 320--330 (2006).
\newblock \doi{10.1057/palgrave.ejis.3000589}

\bibitem{williams2015}
Williams, M.D., Rana, N.P., Dwivedi, Y.K.: The unified theory of acceptance and
  use of technology (utaut): a literature review.
\newblock Journal of Ent Info Management \textbf{28}(3), 443–488 (2015)

\bibitem{yui}
YUI: Yui (2017).
\newblock \urlprefix\url{http://yuilibrary.com}

\end{thebibliography}

\section{Appendix}
\label{appendix}
\subsection*{Interview skeleton}
We developed a generic open-ended interview skeleton, which was maintained to guide the interviewer towards the logical flow of the questions to be asked.

\begin{enumerate}
\item How long have you been developing for the Web?
\item Which was your first programming language for the Web?
\item Which languages do you use for the front-end/back- end?
\item Do you rely on any frameworks for developing? If yes which ones? If no, do you develop in pure JavaScript?
\item When did you start learning Java/JavaScript?
\item Have you ever used JavaScript frameworks? Which ones?
\item Which libraries did you use?
\item Why did you choose this scripting framework?
\item What did you like most about this framework/library?
\item  How long did it take you to learn the framework?
\item  What do you think of the quality of the framework?
\item  Would you recommend this framework to others?
\item  Do you think the framework you used was mature enough? How?
\item  Were there any functionalities that you liked or disliked?
\end{enumerate}

The question draft was used as a guideline for several interviews, except for special cases, such as those where the participants had developed their own JavaScript frameworks or were not using any JavaScript framework but were programming in pure JavaScript language. In these cases they were posed questions like:
\begin{enumerate}
\item Which features similar to other frameworks did you include?
\item How do you decide which new features to integrate into your framework?
\item When did you decide to move from frameworks to pure JavaScript?
\item Would you generally recommend pure JavaScript or frameworks?
\end{enumerate}

\section{Vitae}

\begin{figure}[H]
\includegraphics[keepaspectratio=true,width=4cm]{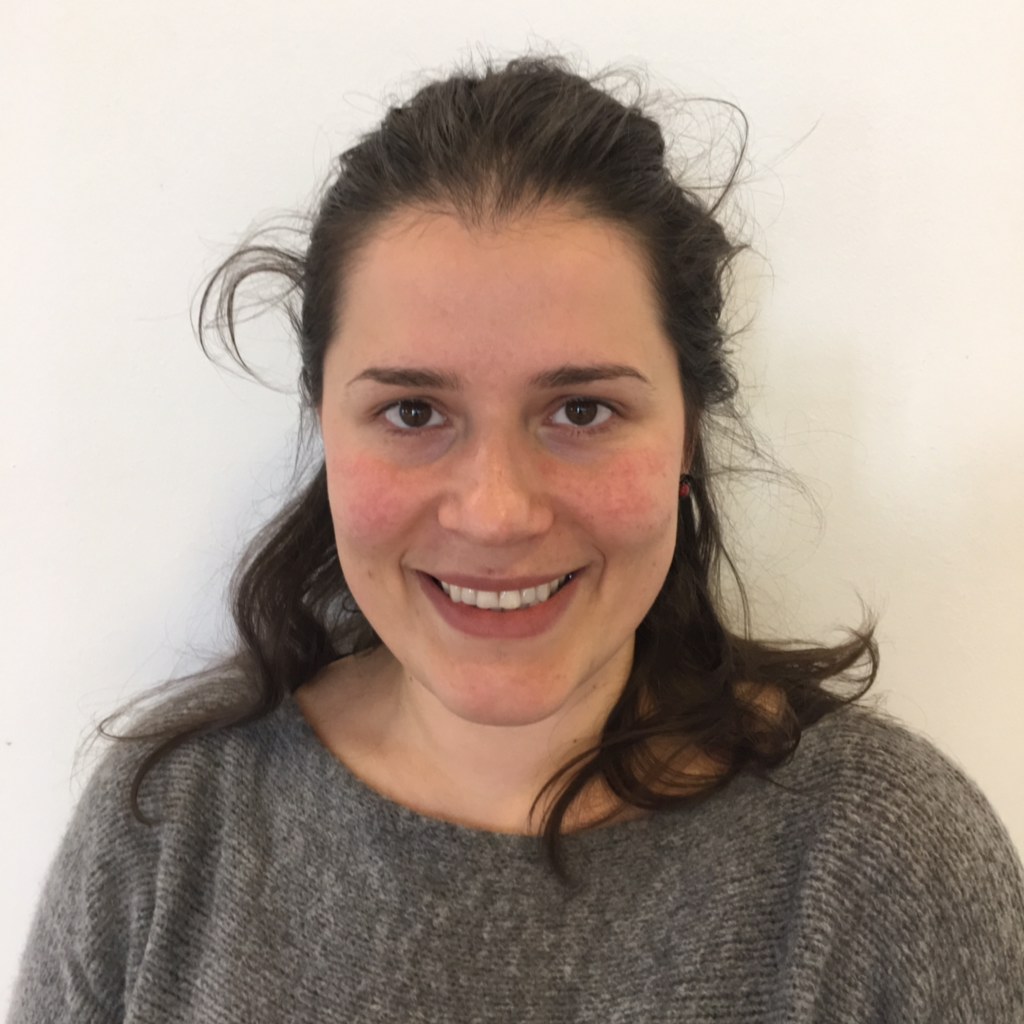}
\end{figure}
\textbf{Amantia Pano} works as a system administrator at the Free University of Bozen-Bolzano in Italy. She holds a MSc degree from the Faculty of Computer Science of the same university. Her research focus is empirical software engineering, in particular on software development process. She collaborates with academic staff and students in preparing the required computational environments for research and teaching activities.

\begin{figure}[H]
\includegraphics[keepaspectratio=true,width=4cm]{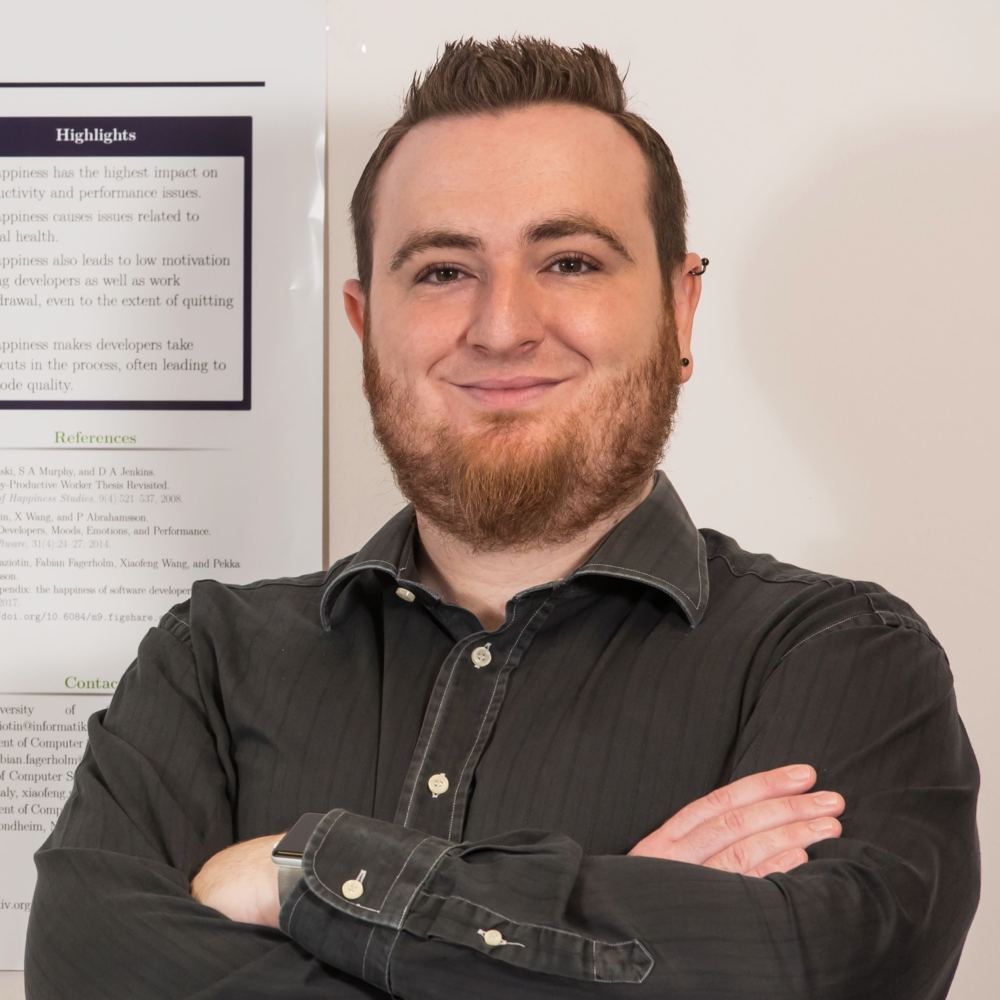}
\end{figure}
\textbf{Daniel Graziotin} is a postdoctoral researcher at the University of Stuttgart, Germany. His research interests include human, behavioral, and psychological aspects of empirical software engineering, studies of science, and open science. He is associate editor at the Journal of Open Research Software and academic editor at the Research Ideas and Outcomes (RIO) journal. Daniel was awarded an Alexander von Humboldt Fellowship for postdoctoral researchers in 2017, the European Design Award (bronze) in 2016, and the Data Journalism Award in 2015. He received his PhD in computer science at the Free University of Bozen-Bolzano, Italy.

\begin{figure}[H]
\includegraphics[keepaspectratio=true,width=4cm]{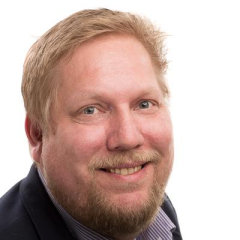}
\end{figure}
\textbf{Pekka Abrahamsson} is professor of Information Systems at the University of Jyväskylä, Finland. Prior to his current position he was a full  professor at NTNU in Norway, dean and full professor at Free University of Bozen-Bolzano, Italy and in University of Helsinki. His research interests are centered on empirical software engineering, agile development and more recently on software startups. He is the recipient  of the Nokia Foundation Award in 2007 for his achievements in the software research. He leads also the SSRN, the global network of software startup researchers. He received his PhD on Software Engineering from University of Oulu.

\end{document}